\documentstyle[epsfig]{aipproc}

\begin{document}
\title{Magnetohydrodynamic Simulations\\ of Black Hole Accretion}

\author{Christopher S. Reynolds$^1$, Philip J. Armitage$^2$ and 
         James Chiang$^{1,3}$}
\address{$^1$JILA, Campus Box 440, University of Colorado, Boulder, CO 80303\\
         $^2$School of Physics and Astronomy, University of St Andrews, Fife, KY16 9SS, UK\\
	 $^3$Laboratory for High Energy Astrophysics, Code 660, NASA/Goddard Space 
	    Flight Center, Greenbelt, MD 20771.}

\maketitle

\begin{abstract}
We discuss the results of three-dimensional magnetohydrodynamic simulations, 
using a pseudo-Newtonian potential, of thin disk $(h/r \approx 0.1)$ accretion 
onto black holes.
We find (i) that magnetic stresses persist within $r_{\rm ms}$, the marginally 
stable orbit, and (ii) that the importance of those stresses for 
the dynamics of the flow depends upon the strength of magnetic 
fields in the disk outside $r_{\rm ms}$. Strong disk magnetic fields 
($\alpha \gtrsim 0.1$)
lead to a gross violation of the zero-torque boundary condition at $r_{\rm ms}$, 
while weaker 
fields ($\alpha \sim 10^{-2}$) produce results more akin to traditional 
models for thin disk accretion onto black holes. Fluctuations in the 
magnetic field strength in the disk could lead to changes in the 
radiative efficiency of the flow on short timescales.
\end{abstract}

\section*{Introduction} 
There is now a consensus that, in well-ionized accretion disks,
magnetorotational instabilities (MRI) create turbulence that provides  
the `anomalous viscosity' required to drive accretion \cite{bh91}.
This removes a major uncertainty that afflicted previous theoretical 
models for black hole accretion, and opens the possibility of 
using numerical simulations to study directly the structure 
and variability of the accretion flow. Questions that we 
might hope to address include:
\begin{itemize}
\item
The magnetohydrodynamics (MHD) of the flow as it crosses $r_{\rm ms}$, 
the marginally stable circular orbit. Recent work has suggested that 
MHD effects interior to $r_{\rm ms}$ (in the `plunging' region) 
could invalidate existing 
models of black hole accretion, with consequences 
that include an increase in the predicted radiative efficiency 
of thin disk accretion \cite{k99,g99,ak00}. This suggestion 
remains controversial \cite{p00}. 
\item
The predicted variability in emission from the inner disk.
\item
The structure of the disk magnetic field and the rate of transport 
of magnetic flux through $r_{\rm ms}$, which together determine 
the efficiency of the Blandford-Znajek mechanism for extracting 
spin energy of the black hole \cite{bz77,ga97,lop99}.
\end{itemize} 
This article presents results from simplified MHD simulations\cite{arc01,ra01}
of black hole accretion that focus on the first of these 
questions. We outline the numerical 
approach, and discuss our results and how they compare 
with those of other groups \cite{h00,hk01,h01}.
Our conclusion from work to date is that MHD 
effects in the plunging region {\em can} have an important 
influence on the dynamics of the inner disk flow, provided 
that the magnetic fields in the disk are already moderately 
strong. 

\section*{Numerical simulations}
Ideally, we would like to simulate a large volume of disk (to allow for 
global effects\cite{a98} and ease worries about treatment of the boundaries),
for a long time period (to average out fluctuations), 
at high resolution (to resolve the most unstable scales of the 
magnetorotational instability), with the most realistic physics 
possible. Unfortunately, we can't, so compromises are 
needed. Our approach has been to simulate the 
simplest disk model that we believe includes the essential 
physical effects, while aiming for the highest resolution near 
and inside the last stable orbit.

\begin{table}
\caption{Summary of the simulations discussed in this article. All 
         runs have the same sound speed, $c_s/v_\phi = 0.065$ (evaluated at
	 $r_{\rm ms}$), 
	 spatial domain $(0.666 < r/r_{\rm ms} < 3.3$, $z/r_{\rm ms} = \pm 0.166$, 
	 $\Delta \phi = 45^\circ)$, and resolution $(n_r = 200$, 
	 $n_z = 40$, $n_\phi = 60)$. The time units are such that the 
	 orbital period at the last stable orbit is $P=7.7$.}
\label{table1}
\center
\begin{tabular}{lllcc}
Run & Vertical boundary & Initial field & Final time & Output times \\
& conditions & & & \\
\hline
Azimuthal field & $B_z=v_z=0$ & $\beta_\phi=100$ & 600 & 400-600 \\
Vertical field & Periodic & $\beta_z=5000$ & 250 & 100-250 \\
High saturation & Periodic & $\beta_z=500$ & 150 & 62.5-150 \\
\hline
\end{tabular}
\end{table}

We use the ZEUS MHD code \cite{sn92a,sn92b} to 
solve the equations of ideal MHD within a 
restricted `wedge' of disk in cylindrical $(z,r,\phi)$ geometry. The 
equation of state is isothermal, and a Paczynski-Wiita potential \cite{pw80} 
is used to model the effect of a last stable orbit within the Newtonian 
hydrocode. To further simplify the problem, we ignore the vertical 
component of gravity and consider an unstratified disk model. The 
boundary conditions are set to outflow at the radial boundaries, are 
periodic in azimuth, and are either periodic or reflecting in $z$.
The simulations begin with a stable, approximately Gaussian 
surface density profile outside $r_{\rm ms}$, which is threaded 
with a weak magnetic field. This initial seed field has a constant 
ratio of thermal to magnetic energy $\beta$, and is either 
vertical or azimuthal. We evolve this setup, which is 
immediately unstable to the MRI, until a significant fraction 
of the mass has been accreted, and plot results from timeslices 
towards the end of the runs when the magnetic fields have reached 
a saturated state.

Table \ref{table1} summarizes the parameters of three simulations, 
which are improved versions of those previously 
reported \cite{arc01}. The simulated disks are `thin' in the 
sense that pressure gradients are negligibly small in the 
disk outside $r_{\rm ms}$. Specifically, the ratio of 
sound speed to orbital velocity, which in a stratified 
disk is approximately 
equal to $h/r$, is $\lesssim 0.1$ at the last stable orbit. We 
have not (yet) attempted the more difficult task of simulating 
very thin disks, which have longer viscous timescales, and 
caution against extrapolating our conclusions into that regime.

Following a suggestion by Charles 
Gammie (personal communication), we investigated whether varying 
the magnetic field strength in the {\em disk} (i.e. at $r > r_{\rm ms}$) 
led to changes in the dynamics of the flow within the plunging 
region. To vary the field strength in the simulations, we 
make use of the fact that the saturation level of the 
MRI can be altered by varying the net flux of seed fields in 
the initial conditions. Local simulations\cite{g98} show that the 
resultant Shakura-Sunyaev $\alpha$ parameter\cite{ss73} is 
approximately,
\begin{equation}
 \alpha \sim 10^{-2} + 4 {{\langle v_{Az} \rangle}  \over c_s} 
 + {1 \over 4} {{\langle v_{A \phi} \rangle}  \over c_s}
\end{equation}
where $v_{Az}$ and $v_{A \phi}$ are the Alfven speeds for 
initial conditions with uniform vertical and azimuthal seed fields, 
and $c_s$ is the sound speed. For our simulations, we find that 
the range of initial conditions (and vertical boundary 
conditions) shown in Table \ref{table1} leads to a variation 
in $\alpha$ between $10^{-2}$ and $10^{-1}$ in the disk.

Of course, this is a numerical trick. The {\em true} value of 
$\alpha$ in the disk immediately outside $r_{\rm ms}$ 
will probably depend upon details of the disk physics (for example,  
the relative importance of gas and radiation pressure), and may vary 
with time.

\section*{Dynamics of the flow crossing $r_{\rm ms}$}

\begin{figure}
\centerline{\epsfig{file=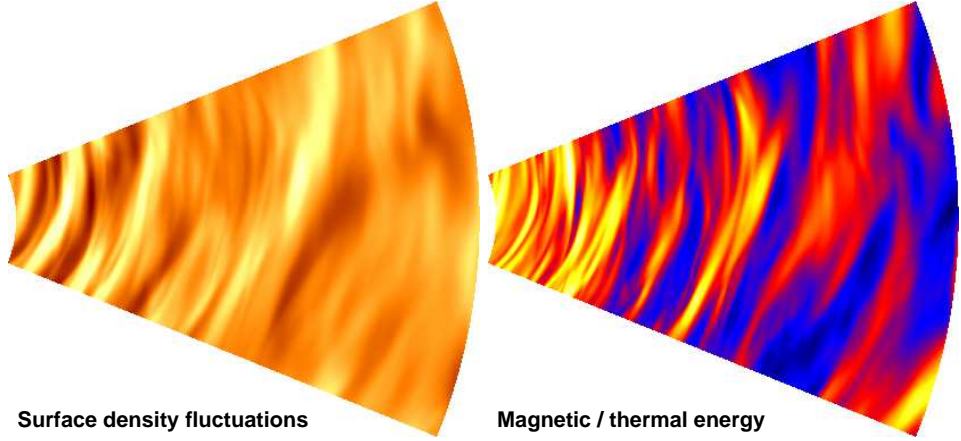,height=2.3in,width=5.0in}}
\vspace{10pt}
\caption{Maps showing (left panel) the surface density fluctuations and 
	 (right panel) the ratio of the energy density in magnetic fields 
         to the thermal energy, from a simulation in which the saturation 
	 value of the magnetorotational instability in the disk was artificially 
	 boosted. The spatial domain covers $0.666 < r/r_{\rm ms} < 3.3$. A 
	 clear increase in the relative importance of 
	 magnetic fields in the inner regions of the disk, and within 
	 the marginally stable orbit, is obtained.}
\label{fig1}
\end{figure}

Figure 1 illustrates the geometry of the simulations. All MHD 
disk simulations look broadly similar, and these are no 
exception. We obtain a pattern of surface density fluctuations 
that are strongly sheared by the differential rotation, and 
disk magnetic fields that are predominantly azimuthal. A map 
of the ratio of magnetic to thermal energy, also shown in 
the Figure, displays clearly the predicted increase\cite{k99} 
in the relative importance of magnetic fields near and interior 
to the marginally stable orbit. Determining the influence 
of these fields upon the dynamics of the flow in the inner 
disk and in the plunging region is the main goal of the 
calculations.

\begin{figure}[t!]
\centerline{\epsfig{file=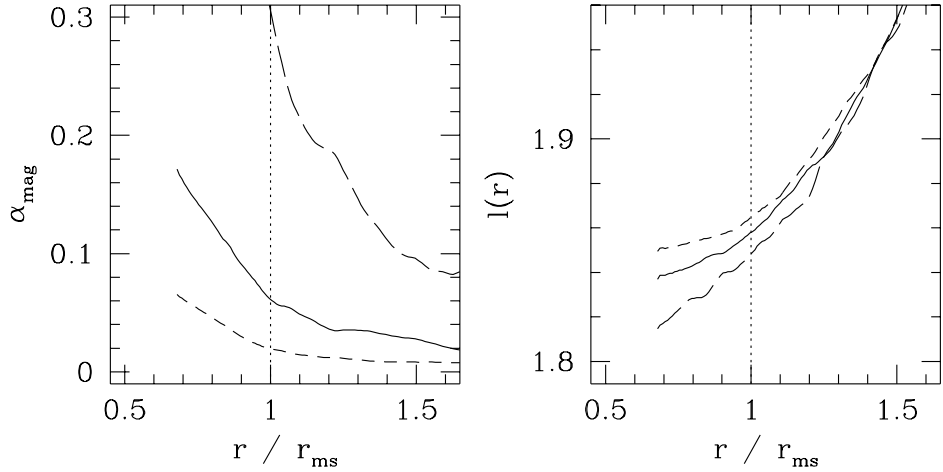,height=5.0in,width=5.0in}}
\vspace{-2.0truein}
\caption{The dynamics of the flow inside the marginally stable 
	orbit is correlated with the magnetic field strength 
	in the disk. The left panel shows the contribution 
	made by magnetic stresses to the Shakura-Sunyaev $\alpha$ 
	parameter as a function of radius for the three runs:
	azimuthal initial field (short dashes), vertical initial field 
	(solid), and strong vertical initial field (long dashes). The 
	right panel shows the resulting specific angular momentum 
	of the flow. All results are averaged over several independent 
	timeslices to reduce fluctuations.
	Strong disk fields lead to a larger violation 
	of the zero-torque boundary condition at $r_{\rm ms}$.}
\label{fig2}
\end{figure}

Figure \ref{fig2} shows the magnetic torque as a function 
of radius in the three simulations, expressed as the 
magnetic contribution to the equivalent Shakura-Sunyaev 
$\alpha$ parameter,
\begin{equation} 
 \alpha_{\rm mag} = {2 \over 3} \langle { {-B_r B_\phi} \over
 {4 \pi \rho c_s^2} } \rangle.
\end{equation}
There are also hydrodynamic (Reynolds) stresses, which are 
somewhat harder to measure\cite{h01}, but which are found 
to be substantially smaller than the magnetic stresses in 
local simulations.  

The three choices for the initial flux and vertical boundary 
conditions lead to large variations in the saturation level 
of the MRI and associated $\alpha_{\rm mag}$. The run with 
an initially azimuthal field produces an  
$\alpha_{\rm mag} \sim 10^{-2}$, while the run with a 
relatively strong initial vertical flux leads to a disk $\alpha_{\rm mag}$ 
that exceeds 0.1. This increase is qualitatively 
in agreement with local simulations\cite{g98}.
For the run with the strongest field, the 
magnetic field energy density in the disk near $r_{\rm ms}$ 
is on average near equipartition with the thermal energy 
$(\beta \sim 1)$. There are large fluctuations with time, 
however, including brief periods where the magnetic energy 
substantially exceeds the thermal energy. In all 
runs, the relative importance of magnetic fields compared 
to the thermal energy increases within $r_{\rm ms}$.

Figure \ref{fig2} also shows how the torques influence the 
dynamics of the flow. We plot the specific angular momentum 
of the flow $l$ as a function of radius. Hydrodynamic models 
for thin disk accretion onto black holes obtain 
$dl/dr = 0$ within the last stable orbit, corresponding to 
a zero-torque boundary condition for the disk\cite{mp82}.
A non-zero $dl/dr$ implies transport of angular momentum 
(and implicitly energy) into the disk from within the 
plunging region. We find that $dl/dr$ in the plunging 
region correlates with $\alpha_{\rm mag}$ in the disk. 
Weak disk fields lead to a small (but significant) 
decline in $l$ within the last stable orbit, while 
strong fields lead to a steeply declining specific angular 
momentum profile at small radii. The former behavior is 
similar to that seen in our earlier simulations\cite{arc01} 
(which also had rather weak fields), while the latter 
is comparable to the results obtained by Hawley\cite{h00} 
and Hawley and Krolik\cite{hk01}. Their global simulations, 
which are substantially more ambitious than ours in terms of the 
included physics and spatial domain, did indeed generate 
relatively strong magnetic fields. 

\section*{Discussion}

The limitations of the current simulations are myriad and obvious. We 
are accutely aware that effects in the disk corona could be 
important\cite{ms00,mhr00}, and that there is more to General 
Relativity than a pseudo-Newtonian potential\cite{kmsk00,f00}. Nonetheless, 
we believe that some conclusions can be drawn from existing work:
\begin{itemize}
\item
Unstratified global simulations confirm that magnetic torques persist 
within the last stable orbit, but suggest that if $\alpha \sim 10^{-2}$ 
their influence on the dynamics 
of the flow is relatively modest\cite{arc01,h01}. By modest we mean 
that the gradient of the specific angular momentum, $dl/dr$, is 
non-zero but small at and inside $r_{\rm ms}$. If these 
simulations reflect reality, existing models of black hole accretion 
would be a pretty good approximation for thin disks\cite{p00}.
\item
A gross violation of the zero-torque boundary condition at $r_{\rm ms}$ 
is also possible\cite{h00,hk01}. This would increase the radiative 
efficiency of thin disk accretion above the usual $\epsilon \approx 0.1$, 
and have other consequences\cite{k99,g99,ak00}. We believe
that these strong effects {\em only occur} if $\alpha \gtrsim 0.1$ in the disk. 
This would be consistent with existing simulations\cite{arc01,h00}, and 
with the results presented here.
\end{itemize}  
For observations, these results suggest that the radiative efficiency 
of thin disk accretion may vary, both between systems, and in an 
individual system if the magnetic field strength in the disk varies 
with time.

\section*{Acknowledgments}

We thank the developers of ZEUS and ZEUS-MP for making these codes 
available as community resources.
P.J.A. thanks JILA for hospitality during the course of part of this work.
C.S.R. acknowledges support from Hubble Fellowship grant HF-01113.01-98A.
This grant was awarded by the Space Telescope Science Institute, operated 
by AURA for NASA under contract NAS 5-26555. C.S.R. also thanks support from  
the NSF under grants AST 98-76887 and AST 95-29170. J.C. was supported 
by NASA/ATP grant NAGS 5-7723.

\end{document}